\documentclass[twocolumn,aps,prb,showpacs,superscriptaddress,longbibliography,10pt]{revtex4-2} 
\usepackage[colorlinks=true,linkcolor=blue,citecolor=blue]{hyperref}
\usepackage{amsfonts}
\usepackage{subfigure}
\usepackage{amsmath}
\usepackage{mathrsfs}
\usepackage{amssymb}
\usepackage{amsbsy}
\usepackage{epsfig}
\usepackage{graphicx}
\usepackage{epstopdf}
\usepackage{mathdots}
\usepackage{color,xcolor}
\usepackage{cleveref}
\usepackage{float}
\usepackage{graphicx}
\usepackage{rotating}
\usepackage{physics}
\usepackage{nicefrac}
\usepackage{soul}
\sethlcolor{yellow}
\setstcolor{blue}
\setulcolor{blue}
\usepackage[colorlinks=true, linkcolor=blue]{hyperref}
\usepackage{cleveref}
\usepackage{soul}
\soulregister\ref7 
\soulregister\citealt7 
\usepackage{cancel}

\begin{document}
\title{Quasiperiodic Skin Criticality in an Exactly Solvable Non-Hermitian Quasicrystal}

\author{Zhangyuan Chen}
\affiliation{Department of Physics, Jiangsu University, Zhenjiang, 212013, China}

\author{Muhammad Idrees} 
\affiliation{Department of Physics, Jiangsu University, Zhenjiang, 212013, China}

\author{Ying Yang} 
\affiliation{Department of Physics, Jiangsu University, Zhenjiang, 212013, China}

\author{Xianqi Tong} \altaffiliation{xqtong@ujs.edu.cn}
\affiliation{Department of Physics, Jiangsu University, Zhenjiang, 212013, China}

\author{Xiaosen Yang} \altaffiliation{yangxs@ujs.edu.cn} \affiliation{Department of Physics, Jiangsu University, Zhenjiang, 212013, China}

\date{\today}

\begin{abstract}
Critical states in quasiperiodic systems defy the conventional dichotomy between extended and localized states. In this work, we demonstrate that non-Hermiticity fundamentally reshapes this paradigm by giving rise to an exactly solvable quasiperiodic critical phase with no energy selectivity. We introduce a non-Hermitian quasiperiodic lattice based on a modulated Hatano-Nelson model and uncover a new universality class of quasiperiodic skin criticality, in which all eigenstates share an identical multifractal spatial structure. Through a nonunitary gauge transformation, the system is mapped onto a disorder-free lattice, enabling exact analytical solutions for the full spectrum and eigenstates. As a consequence, the inverse participation ratio is strictly energy-independent and controlled solely by a global phase. We further show that this criticality persists in multiband lattices, establishing a general and analytically controlled framework for non-Hermitian quasiperiodic critical phenomena.
\end{abstract}

\maketitle
\section{Introduction}
\begin{figure}[t]
    \centering
    \includegraphics[width=1\linewidth]{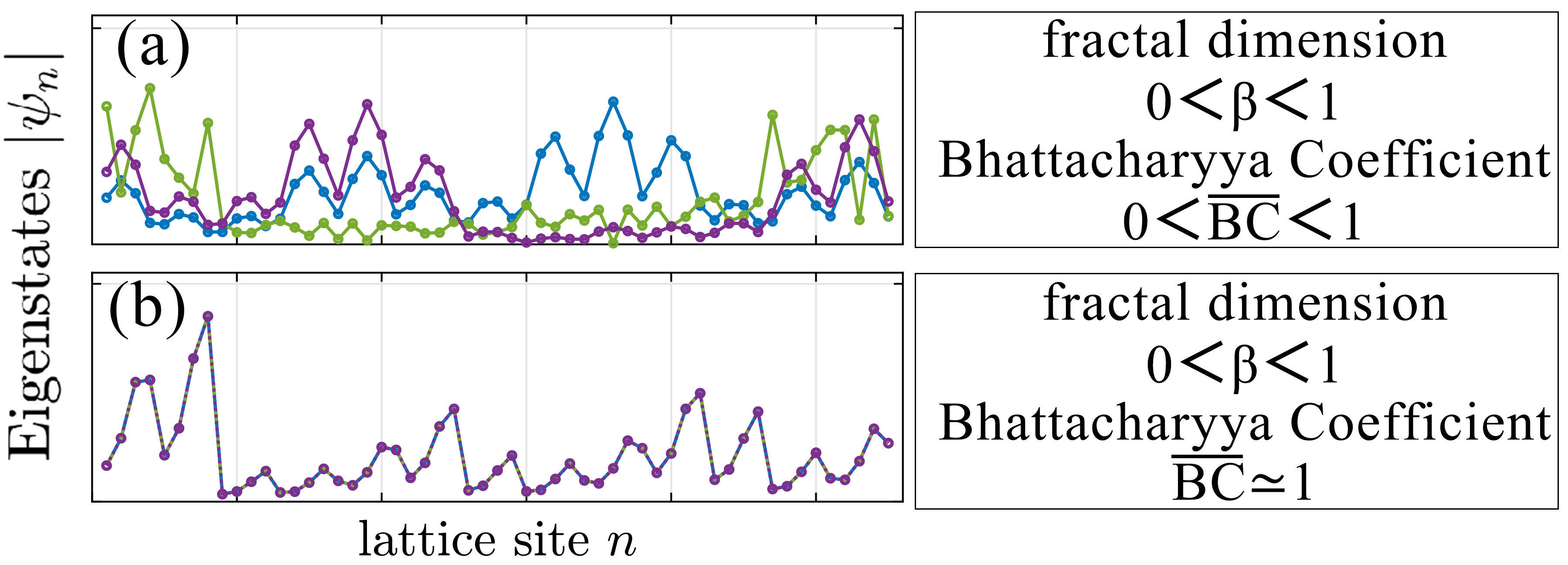}
    \caption{Comparison between conventional quasiperiodic criticality and the QNHSC. The panels illustrate the spatial profiles of three representative eigenstates, $|\psi_n|$ (colored lines), emphasizing the distinctive criticality of these two phases.
    (a) Conventional quasiperiodic criticality: Eigenstates display energy-dependent multifractal fluctuations. Different eigenstates are spatially separated, leading to a partial spatial overlap quantified by $\overline{\mathrm{BC}}<1$.
    (b) QNHSC: All eigenstates exhibit the same criticality. This collective condensation results in an almost perfect spatial overlap, characterized by $\overline{\mathrm{BC}}\simeq1$.}
    \label{figure1}
\end{figure}

Conventional band theory predicts extended Bloch states in periodic potentials~\cite{Bloch1929Quantenmechanik,grosso2013solid,kittel2018introduction}, whereas Anderson localization establishes that disorder induces exponential localization~\cite{Anderson1958Absence,Ishii1973Localization,THOULESS1974Electrons,Abrahams1979Scaling,Evers2008Anderson,Das2023Absence}. Quasiperiodic systems bridge this dichotomy~\cite{Hatsugai1990Energy, jaric2012introduction}. Lacking translational invariance yet possessing long-range order, they host critical states characterized by scale-invariant multifractality~\cite{Kohmoto1987Critical, Kohmoto1987Localization, Vladimir2009Optical, Deguchi2012QuantumCritical}. Such states defy the standard extended--localized classification~\cite{Yang2017Dynamical,goblot2020emergence,shimasaki2024anomalous,Dotti2025Measuring}, offering a unique platform to investigate critical phenomena~\cite{Yao2019Critical,Wang2020Realization,Wang2021ManyBody, Xiao2021Observation,Tong2021Dynamics, Gon2023Critical,Yang2024Exploring,Duncan2024Critical,Yao2024Wave,Zhang2025Critical}.

Non-Hermiticity introduces novel phenomena~\cite{Bender1998Real,Zeuner2015Observation,Leykam2017Edge,Yao2018Edge,Lee2019Topological,Ashida2020NonHermitian,Zhang2020Correspondence,li2020critical,Bergholtz2021Exceptional,zhang2021observation,Li2022Topological,zhang2022universal,jiGeneralized2024, FuPRBbraiding, Li2025Exact,Zhang2025Yang-Lee,Wang2025Theory} distinct from their Hermitian counterparts, most notably the non-Hermitian skin effect (NHSE), which has been extensively explored in theory~\cite{Okugawa2020Second-order,Yokomizo2021Scaling,Liang2022Dynamic,lin2023topological,Li2024DynamicNHSE,Yoshida2024Non-Hermitian,Ma2024Non-Hermitian,yang2026kagome} and experiment~\cite{weidemann2020topological,Gu2022TransientNHSE, Yiling2022Flexible,zhou2023observation, Lin2024Observation,Wang2025Nonlinear, Wang2025Tunable,Wang2026One-Dimensional}. Unlike Hermitian systems, where bulk states are extended, the bulk states of non-Hermitian systems can exhibit a macroscopic accumulation at the boundaries\cite{Okuma2020Topological}. The interplay between quasiperiodicity and non-Hermiticity has unveiled rich phenomena~\cite{Jiang2019Interplay, Junmo2023Localization, Shi2024Delocalization, Zhou2024Entanglement, Rangi2024Engineering, Zheng2025Emergent, Li2025Anderson-skin, Gandhi2025Superconducting, liang2025size}, including complex mobility edges~\cite{Lin2022Topological, Wang2024Exact, Li2024Ring} and topology dependent localization transitions~\cite{Tang2021Localization, Longhi2021Phase,Zhou2023Non-Abelian, Wang2025Quasiperiodicity}. Although these studies indicate that non-Hermiticity qualitatively modifies quasiperiodic localization, rigorous analytical frameworks remain scarce. In particular, the field lacks exactly solvable models that can clearly capture the intrinsic nature of critical states in this regime. Consequently, fundamental questions remain: Can non-Hermiticity induce new universality classes of criticality? 

In this work, we explore this question by introducing an exactly solvable non-Hermitian quasicrystal derived from a modulated Hatano–Nelson (HN) model. Unlike conventional quasiperiodic criticality, which typically exhibits multifractal intensity fluctuations [Fig.~\ref{figure1}(a)], we uncover a novel phase, termed quasiperiodic non-Hermitian skin criticality (QNHSC), where all eigenstates exhibit an identical, energy-independent multifractal structure controlled by a global phase [Fig.~\ref{figure1}(b)]. By employing a nonunitary gauge transformation, we map the system onto a disorder-free lattice, enabling the derivation of exact analytical expressions for the eigenstates and inverse participation ratio ($\mathrm{IPR}$). This analytical establishes a rigorous framework for understanding the interplay of non-Hermiticity, quasiperiodicity, and criticality. Furthermore, we demonstrate that this critical behavior persists in multiband lattices, confirming the generality of the phenomenon beyond single-band models.

\section{Hatano--Nelson model}
\begin{figure}[t]
    \centering
    \includegraphics[width=1\linewidth]{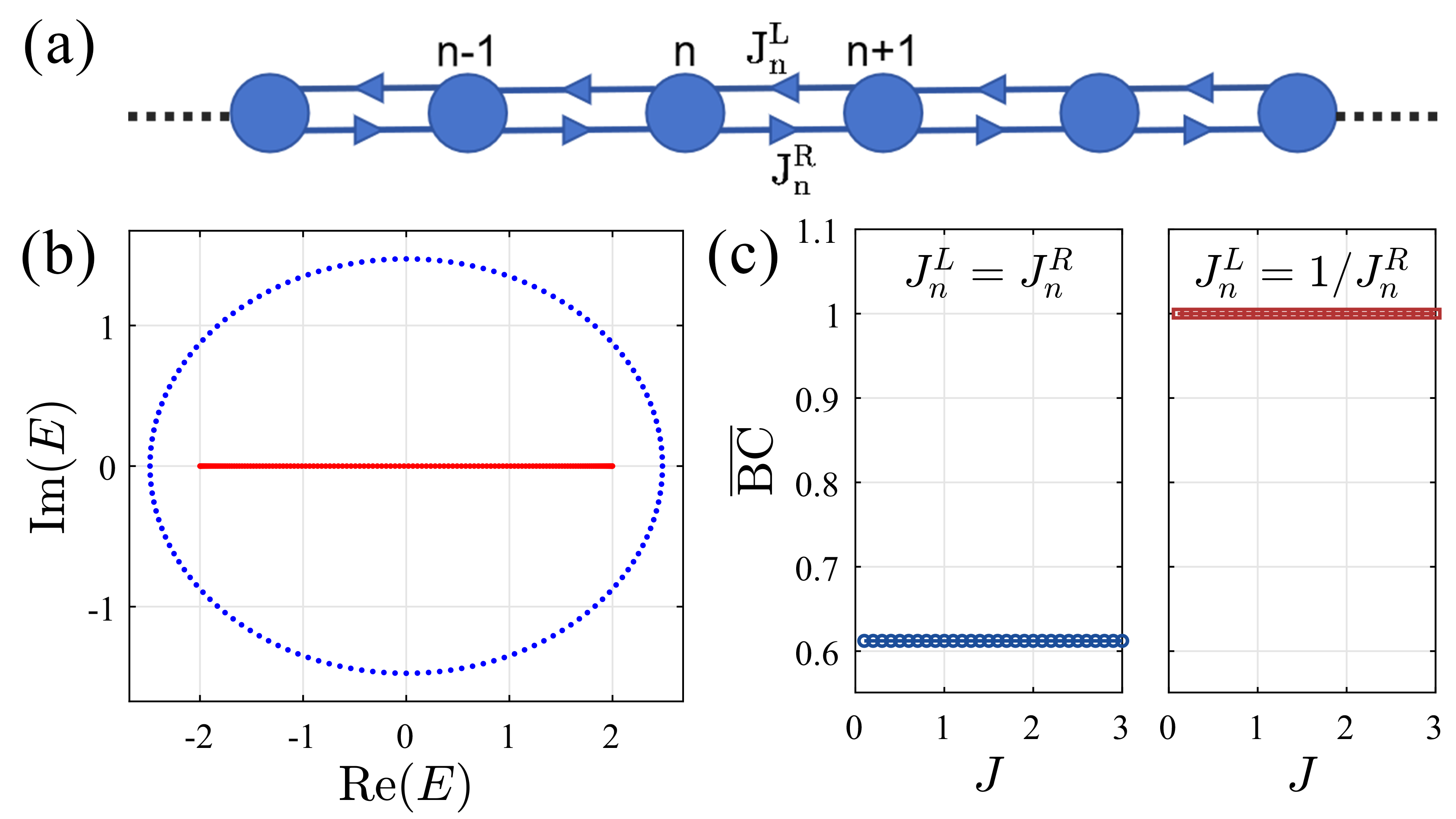}
    \caption{(a) Schematic illustration of the quasiperiodically modulated HN lattice.
    (b) Complex energy spectrum plotted in the complex plane under PBC (blue points) and OBC (red points). Parameters are $N=89$, $J=1$, and $\theta=0$. 
    (c) The $\overline{\mathrm{BC}}$ versus quasiperiodic modulation strength $J$ for system size $N=2584$ and global phase $\theta=0$. Left: Hermitian quasiperiodic model with $J_n^L=J_n^R$. Right: Non-Hermitian quasiperiodic model with $J_n^L = 1/J_n^R$.}
    \label{figure2}
\end{figure}

The HN model serves as a paradigmatic platform in non-Hermitian physics~\cite{Hatano1996Localization, Hatano1997Vortex, Hatano1998Non-Hermitian}, providing a minimal and analytically transparent setting for understanding the NHSE and its associated spectral topology. Originating from intrinsic nonreciprocity, this model has been instrumental in demonstrating how boundary conditions can fundamentally reshape the bulk spectrum~\cite{Guo2021Exact,Longhi2021Spectral}. To explore the emergence of QNHSC, we consider a HN chain subject to quasiperiodic modulation~\cite{Midya_Topological2024}. The resulting Hamiltonian, schematically illustrated in Fig.~\ref{figure2}(a), reads
\begin{equation}
\hat{H}=\sum_{n=1}^{N-1}\left(J_{n}^{R} \hat{c}_{n+1}^{\dagger} \hat{c}_{n}+J_{n}^{L} \hat{c}_{n}^{\dagger} \hat{c}_{n+1}\right)+\hat{H}_{B},
\label{HNmodel}
\end{equation}
where $N$ denotes the system size, $\hat{c}_{n}^{\dagger}$ ($\hat{c}_{n}$) creates (annihilates) a particle at lattice site $n$, and $J_{n}^{R}$ ($J_{n}^{L}$) represents the right (left) hopping amplitude. $\hat{H}_{B}$ is the Hamiltonian term that specifies the lattice boundary conditions: $\hat{H}_{B}=0$ corresponds to open boundary conditions (OBC), whereas for periodic boundary conditions (PBC), $\hat{H}_{B} = J_{N}^{R} \hat{c}_{1}^{\dagger} \hat{c}_{N} + J_{N}^{L} \hat{c}_{N}^{\dagger} \hat{c}_{1}$. Quasiperiodicity is introduced by specifying the site-dependent hopping amplitudes as:
\begin{equation}
J_{n}^{R}=J \cos [2 \pi \alpha(n+1 / 2)+\theta], \quad J_{n}^{L}=1/J_{n}^{R},
\label{Quasiperiodic_Hopping_Amplitudes}
\end{equation}
where $J$ sets the strength of the quasiperiodic modulation. The parameter $\theta$ sets a global phase of the quasiperiodic modulation. Although varying $\theta$ corresponds to a spatial translation of a deterministic potential, it uniquely determines the spatial structure of the eigenstates in the present model. The quasiperiodicity is governed by the irrational modulation frequency $\alpha$, chosen as the inverse golden mean $\alpha=(\sqrt{5}-1)/2$. In numerical calculations, this limit is realized by using a sequence of rational approximants $\alpha \simeq F_{n}/F_{n+1}$, where $F_n$ denotes the $n$th Fibonacci number.

The single-particle eigenstates $\psi_{n}$ and the corresponding eigenenergies $E$ satisfy the tight-binding difference equation
\begin{equation}
E \psi_{n}=J_{n}^{L} \psi_{n+1}+J_{n-1}^{R} \psi_{n-1},
\label{The_Spectral_Problem}
\end{equation}
with the boundary conditions $\psi_{0}=\psi_{N+1}=0$ under OBC or $\psi_{n+N}=\psi_{n}$ under PBC. A key structural ingredient of the model is the reciprocal constraint $J_n^L = 1/J_n^R$, which is not a fine-tuned parameter choice but an intrinsic feature of the construction. As shown below, this constraint enables an exact mapping to a solvable model via a nonunitary gauge transformation.

In the periodic limit ($\alpha=0$), the Hamiltonian reduces to the standard HN model exhibiting the NHSE~\cite{Claes2021Skin,Zhang2023BulkBoundary,Midya_Topological2024}. To explicitly solve the quasiperiodic case ($\alpha=(\sqrt{5}-1)/2$), we introduce the following nonunitary gauge transformation:
\begin{equation}
\psi_n = \phi_n \prod_{l=1}^{n-1} J \cos[2\pi\alpha (l + 1/2) + \theta].
\label{Nonunitary_Gauge_Transformation}
\end{equation}
This transformation corresponds to a local rescaling of the eigenstates amplitudes. Owing to the nonreciprocal relation $J_n^L = 1/J_n^R$, this mapping simultaneously eliminates the non-Hermitian nonreciprocity and the spatial modulation of the hopping terms. This simplification allows for a direct analytical solution of the spectrum. 

Substituting the transformation into Eq.~\eqref{The_Spectral_Problem}, the model maps onto the eigenvalue problem of a disorder-free one-dimensional tight-binding chain: $E \phi_{n}= \phi_{n+1}+\phi_{n-1}$. 
The boundary conditions transform accordingly: $\phi_{0} = \phi_{N+1} = 0$ for OBC, and $\phi_{N+1} = \phi_{1} e^{-\Gamma_{N+1}}$ for PBC, where $\Gamma_{N+1} \equiv \sum_{l=1}^{N} \ln\!\left|J \cos\!\left(2\pi\alpha\left(l+\tfrac{1}{2}\right)+\theta\right)\right|$. For OBC, the solutions are standing states with a purely real spectrum:
\begin{equation}
\begin{aligned}
&\phi_n = \sin(qn), \quad E_{\text{OBC}} = 2\cos q, \\
&q = \frac{m\pi}{N+1} \quad (m = 1, 2, \dots, N).
\end{aligned}
\label{OBC_Solution}
\end{equation}
For PBC, the eigenstates take the form of Bloch states with a complex wave number, leading to a complex spectrum:
\begin{equation}
\begin{aligned}
&\phi_n = e^{i q n - \frac{n}{N} \Gamma_{N+1}}, \quad E_{\text{PBC}} = 2\cos(q+ \frac{i}{N} \Gamma_{N+1}),\\
&q = \frac{2\pi m}{N} \quad (m = 0,1,\dots,N-1).
\end{aligned}
\label{PBCSpectral_Problem_Solution}
\end{equation}
Thus, the OBC spectrum is entirely real and confined to $[-2,2]$, whereas under PBC the spectrum traces a closed loop in the complex plane, as shown in Fig.~\ref{figure2}(b). This extreme sensitivity of the spectrum to boundary conditions directly signals the NHSE.

\section{Eigenstate characteristics}

\begin{figure*}[t]
    \centering
    \includegraphics[width=0.99\textwidth]{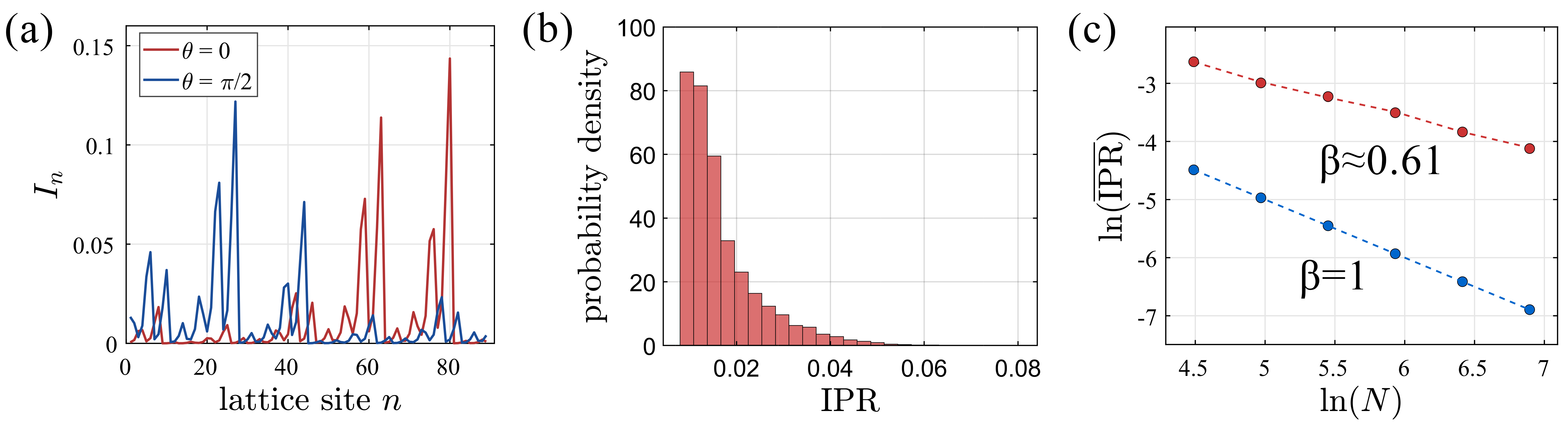}
    \caption{Analysis of eigenstate similarity and scaling behavior in non-Hermitian quasiperiodic HN chains.
    (a) The spatial distributions of PBC eigenstates $I_n$ for two values of the global phase: $\theta=0$ (red) and $\theta=\pi/2$ (blue).
    (b) Probability density function of the IPR values for $N=987$, collected over $10^4$ realizations of the global phase $\theta$. 
    (c) Finite-size scaling of the $\overline{\mathrm{IPR}}$ versus system size $N$ on a logarithmic scale. Red points indicate numerical results, which agree with the theoretical prediction (red dashed line). For comparison, the blue points show the $\overline{\mathrm{IPR}}$ in a disorder-free lattice for $\alpha=0$. $\beta$ is the corresponding fractal dimension.}
    \label{figure3}
\end{figure*}

To characterize the spatial distribution of the PBC eigenstates, we calculate the averaged probability density~\cite{zhang2022universal}:
$I_{n} = \frac{1}{N} \sum_{j=1}^N \left| \psi_{n}^{(j)} \right|^{2}$.  
A hallmark of these eigenstates is that the spatial profile is independent of eigenenergy in that the positions of the main and satellite peaks in the lattice is the same for all eigenstates. To quantify this spatial similarity, we compute the averaged Bhattacharyya Coefficient ($\overline{\mathrm{BC}}$)~\cite{Bhattacharyya1943Measure}:
\begin{equation}
\overline{\mathrm{BC}}=\frac{1}{N} \sum_{j=1}^{N}\left(\sum_{n=1}^{N} \sqrt{I_{n} \cdot\left|\psi_{n}^{(j)}\right|^{2}}\right).
\label{BC}
\end{equation}
Values of $\overline{\mathrm{BC}}\simeq1$ indicate nearly perfect spatial overlap, whereas smaller values correspond to distinct eigenstate profiles. As shown in the right panel of Fig.~\ref{figure2}(c), the $\overline{\mathrm{BC}}$ for the non-Hermitian quasiperiodic model remains unity within numerical precision, confirming the exact coincidence of all eigenstate profiles. This behavior contrasts sharply with the Hermitian quasiperiodic model shown in the left panel of Fig.~\ref{figure2}(c), where eigenstates exhibit spatially distinct multifractal patterns ($\overline{\mathrm{BC}}<1$). 
The spatial profile of the eigenstate is not random but depends on the global phase $\theta$, as shown as illustrative examples in Fig.~\ref{figure3}(a). Varying $\theta$ shifts the modulation peaks, highlighting the global phase control over the eigenstate structure. Unlike conventional extended states with uniform density (see Appendix~\ref{Conventional} for details), these states mark the emergence of the QNHSC, exhibiting a collective multifractal structure.

Substituting the nonunitary gauge transformation of Eq.~\eqref{Nonunitary_Gauge_Transformation} into the PBC solution of Eq.~\eqref{PBCSpectral_Problem_Solution}, we obtain an exact analytical expression for the eigenstates:
\begin{equation}
\left| \psi_{n} \right| \propto \exp\left\{\Gamma_{n}-\frac{n}{N} \Gamma_{N+1}\right\},
\label{Analytical_Solution1}
\end{equation}
where the cumulative modulation phase is defined as $\Gamma_n \equiv \sum_{l=1}^{n-1} \ln\!\left|J \cos\!\left(2\pi\alpha\left(l+\tfrac{1}{2}\right)+\theta\right)\right|$ with $\Gamma_1 = 0$. 
Crucially, because the eigenenergy $E$ appears only through the Bloch phase factor $e^{iqn}$ in $\phi_n$, taking the modulus removes all energy dependence. As a result, every eigenstate in the spectrum exhibits an identical multifractal spatial structure, uniquely determined by the quasiperiodic modulation pattern and fully independent of eigenenergy.

The critical nature of the eigenstates is rigorously established by computing the IPR and fractal dimension $\beta$~\cite{Bauer1990Correlation,Evers2000Fluctuations}.
For normalized eigenstates, the IPR and fractal dimension $\beta$ are defined by $\operatorname{IPR} =\sum_{n=1}^{N}|\psi_n|^4$ and $\beta=\lim _{N \rightarrow \infty}[\ln \operatorname{IPR} / \ln (1 / N)]$. In contrast to numerical approaches that extract critical behavior via finite-size scaling, the exact solution provides a direct and fully explicit characterization of the eigenstate structure. We find that the IPR obeys a universal power-law scaling $\operatorname{IPR} \sim N^{-\beta}$ ($0 < \beta < 1$), indicating that the eigenstates are neither extended ($\beta=1$) nor exponentially localized ($\beta=0$), but intrinsically multifractal. For finite system sizes $N$, the IPR is not sample independent, due to the lack of translational invariance, it fluctuates sensitively with the choice of global phase $\theta$. The probability distribution of the IPR values is shown in Fig~\ref{figure3}(b) obtained by varying $\theta$ for a fixed system size $N=987$. Although the IPR varies across different $\theta$ realizations, it is identical for all eigenstates within a given realization, demonstrating that the QNHSC originates from the underlying nonreciprocal configuration, rather than a random spectral property.

To extract the fractal dimension, we perform a finite-size scaling analysis of the averaged IPR $\overline{\mathrm{IPR}} = \langle \operatorname{IPR} \rangle_{\theta},$ defined as the ensemble average over realizations of the global phase $\theta$. Figure~\ref{figure3}(c) shows the size scaling behavior of the $\overline{\mathrm{IPR}}$. The numerical data (red points) display a robust power-law decay, yielding a fractal dimension $\beta \approx 0.61$, which unequivocally confirms the critical nature of the eigenstates. Furthermore, using the exact analytical of eigenstates, we derive the expression of the $\mathrm{IPR}$ (see Appendix~\ref{IPRofHN} for details):
\begin{equation}
\operatorname{IPR}=\frac{\sum_{n=1}^{N} \exp \left(4 \Gamma_{n}-4 \frac{n}{N} \Gamma_{N+1} \right)}{\left[\sum_{n=1}^{N} \exp \left(2 \Gamma_{n}-2 \frac{n}{N} \Gamma_{N+1} \right)\right]^{2}}.
\label{IPR_Formula}
\end{equation}
The theoretical prediction from Eq.~\eqref{IPR_Formula} (red dashed line) exhibits excellent agreement with the numerical data. For comparison, the behavior of the $\overline{\mathrm{IPR}}$ for a disorder-free lattice ($\alpha=0$) is shown by blue points, following the trivial scaling characteristic of extended Bloch states ($\beta=1$). This stark contrast underscores the fundamental distinction between extended states and the QNHSC derived here.

\section{Hatano--Nelson Ladder Model}

\begin{figure}[t]
    \centering
    \includegraphics[width=1\linewidth]{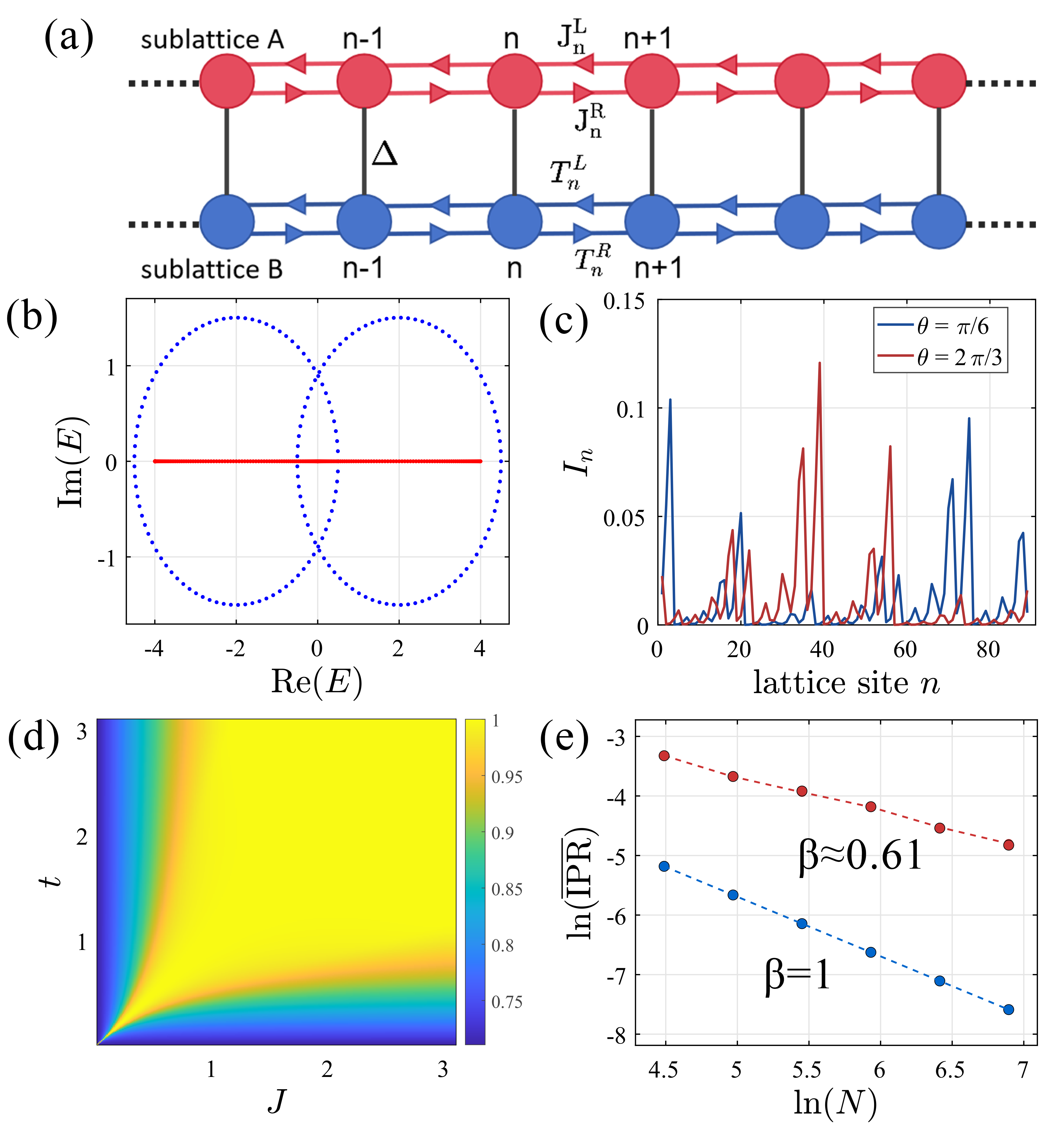}
    \caption{Properties of the non-Hermitian quasiperiodic HN ladder model. 
    (a) Schematic of the HN ladder model with quasiperiodic hopping amplitudes. 
    (b) Energy spectrum under PBC (blue points) and OBC (red points) for $j=t=1$, $\Delta=2$, and system size $N=89$. 
    (c) Representative eigenstate distributions $I_n$ for two distinct realizations of the global phase $\theta$. 
    (d) The $\overline{\mathrm{BC}}$ as a function of modulation $J$ and $t$, with $N=89$, $\Delta=1$, and $\theta=0$. 
    (e) Finite-size scaling of the $\overline{\mathrm{IPR}}$ versus system size $N$ (red points), obtained from $10^3$ realizations of $\theta$. The dashed line shows the theoretical prediction from the analytical IPR expression, while blue points correspond to the disorder-free lattice ($\alpha=0$).}
    \label{figure4}
\end{figure}

To demonstrate the universality of QNHSC beyond single-band models, we construct a HN ladder model, as illustrated in Fig.~\ref{figure4}(a). The Hamiltonian reads:
\begin{equation}
\begin{aligned}
\hat{H} = &\sum_{n=1}^{N-1}\Big(J_{n}^{R}\hat{a}_{n+1}^{\dagger}\hat{a}_{n}+ J_{n}^{L}\hat{a}_{n}^{\dagger}\hat{a}_{n+1}\Big) \\
&+\sum_{n=1}^{N-1}\Big(T_{n}^{R}\hat{b}_{n+1}^{\dagger}\hat{b}_{n}+ T_{n}^{L}\hat{b}_{n}^{\dagger}\hat{b}_{n+1}\Big) \\
&+\Delta\sum_{n=1}^{N}\left(\hat{a}_{n}^{\dagger}\hat{b}_{n}+\hat{b}_{n}^{\dagger}\hat{a}_{n}\right)+ \hat{H}_{B}.
\end{aligned}
\label{HNL}
\end{equation}
To retain exact analytical tractability, we impose inverse hopping relations analogous to the single-chain case:
\begin{equation}
\begin{aligned}
J_{n}^{R} &= J \cos[2\pi\alpha(n+1/2)+\theta], \quad &J_{n}^{L} = 1/J_{n}^{R}, \\
T_{n}^{R} &= t \cos[2\pi\alpha(n+1/2)+\theta], \quad &T_{n}^{L} = 1/T_{n}^{R}.
\end{aligned}
\end{equation}
Here, $\hat{a}_{n}^{\dagger}$ ($\hat{b}_{n}^{\dagger}$) creates a particle on the upper (lower) chain, $J$ and $t$ set the modulation strengths, and $\Delta$ denotes the inter-chain coupling. Eigenvectors are written as $|\psi\rangle=(\psi_1^A,\psi_1^B,\psi_2^A,\psi_2^B,\dots)^T$, leading to the spectral equations:
\begin{equation}
\begin{aligned}
J_{n-1}^{R}\psi_{n-1}^A + J_n^L \psi_{n+1}^A + \Delta \psi_n^B &= E \psi_n^A,\\
T_{n-1}^{R}\psi_{n-1}^B + T_n^L \psi_{n+1}^B + \Delta \psi_n^A &= E \psi_n^B.
\label{Spectral_Problem2}
\end{aligned}
\end{equation}
Upon introducing the nonunitary gauge transformation,
\begin{equation}
\begin{aligned}
\psi_n^A &= \phi_n^A \prod_{l=1}^{n-1} J \cos[2\pi\alpha(l+1/2)+\theta],\\
\psi_n^B &= \phi_n^B \prod_{l=1}^{n-1} t \cos[2\pi\alpha(l+1/2)+\theta].
\label{Nonunitary_Gauge_Transformation2}
\end{aligned}
\end{equation}
Eq.~\eqref{Spectral_Problem2} is transformed into a form with uniform hopping and an effective inter-chain potential:
\begin{equation}
\begin{aligned}
E\phi_n^A &= \phi_{n-1}^A + \phi_{n+1}^A + \Delta \left(\frac{t}{J}\right)^{n-1} \phi_n^B,\\
E\phi_n^B &= \phi_{n-1}^B + \phi_{n+1}^B + \Delta \left(\frac{J}{t}\right)^{n-1} \phi_n^A.
\label{emi_spectral_problem2}
\end{aligned}
\end{equation}
The NHSE manifests in the spectral topology shown in Fig.~\ref{figure4}(b), where the complex PBC loops differ drastically from the collapsed real OBC spectrum. In the generic asymmetric case ($J \neq t$), the nonunitary gauge transformation induces spatially varying inter-chain coupling $\Delta(t/J)^{n-1}$ and $\Delta(J/t)^{n-1}$, which act as effective potentials and break exact translational invariance, leading to energy-dependent localization. In the symmetric case $J = t$, the induced couplings become uniform, restoring universality. The energy-independence of the spatial profile is recovered: all eigenstates share identical amplitudes regardless of band index or energy. This behavior is quantitatively captured by the $\overline{\mathrm{BC}}$. As shown in Fig.~\ref{figure4}(d), the $\overline{\mathrm{BC}}$ reaches unity along the line $J = t$, indicating complete spatial overlap of all eigenstates. Away from this line, the $\overline{\mathrm{BC}}$ is suppressed, signaling the re-emergence of energy-dependent criticality.

When $J = t$, the exact amplitude distribution for both sublattices coincides with the single-chain solution:
\begin{equation}
|\psi_n^A| = |\psi_n^B| \propto \exp\Big\{\Gamma_n - \frac{n}{N} \Gamma_{N+1}\Big\}.
\label{two_band_solution}
\end{equation}
The universal QNHSC is further characterized by the $\overline{\mathrm{IPR}}$. Figure~\ref{figure4}(e) shows the scaling of $\overline{\mathrm{IPR}}$ versus system size $N$, with the average taken over $10^3$ realizations of $\theta$. Based on the exact analytical expression for the eigenstates given in Eq.~\eqref{two_band_solution}, we obtain (see Appendix~\ref{IPRofHNL} for details):
\begin{equation}
\operatorname{IPR}=\frac{\sum_{n=1}^{N} \exp \left(4 \Gamma_{n}-4 \frac{n}{N} \Gamma_{N+1} \right)}{2\left[\sum_{n=1}^{N} \exp \left(2 \Gamma_{n}-2 \frac{n}{N} \Gamma_{N+1} \right)\right]^{2}},
\label{IPR_2band}
\end{equation}
with $\Gamma_n$ the cumulative modulation phase defined previously. The theoretical prediction (red dashed line) shows excellent agreement with the numerical data and yields a fractal dimension $\beta \approx 0.61$, confirming the robustness of the QNHSC in the multiband lattices.

In Fig.~\ref{figure4}(c), we plot the spatial distribution of the eigenstates $I_n$ for two distinct realizations of $\theta$. This behavior is in sharp contrast to that observed in the disorder-free lattice. As shown by the blue points in Fig.~\ref{figure4}(e), the behavior of the $\overline{\mathrm{IPR}}$ for a disorder-free lattice follows the trivial scaling characteristic of extended states ($\beta=1$). The appearance of a fractional dimension $\beta \approx 0.61$ in the non-Hermitian quasiperiodic model marks a transition from trivial extended states to nontrivial QNHSC with intrinsic multifractal properties.

\section{Conclusion}
We identified and analytically characterized a distinct phase of matter, QNHSC, within a non-Hermitian quasiperiodic lattice. Through a rigorous nonunitary gauge transformation, we map the quasiperiodic problem onto an exactly solvable disorder-free system, yielding exact analytical solutions for the critical eigenstates. This analytical control reveals a hallmark signature of the phase: energy-independent multifractality, strictly decoupled from spectral properties. This stands in sharp contrast to energy-dependent behavior typically found in conventional quasiperiodic systems, instead, the spatial profile of the critical states is deterministically dictated by the global modulation phase. We further demonstrate the robustness of this mechanism in a HN ladder model. Our results establish a rigorous analytical benchmark for non-Hermitian quasiperiodic systems and reveal a new universality class of skin-induced critical states, thereby offering a controlled route for exploring these phenomena in engineered photonic\cite{Gong2018Topological,Ozawa2019Topological,Bergholtz2021Exceptional,yang2026kagome}, acoustic\cite{Zhang2019Non-Hermitian, Gao2021NHHigherOrder, Hu2023Anti, Huang2024AcousticResonances}, and cold-atom platforms~\cite{Li2020Topological, Liu2024Complete,Luo2024Quantum}.

\section{Acknowledgments}
This work was supported by Natural Science Foundation of Jiangsu Province (Grant No. BK20231320).

\appendix
\section{Conventional quasiperiodic criticality}
\label{Conventional}
\begin{figure}[t]
    \centering
    \includegraphics[width=1\linewidth]{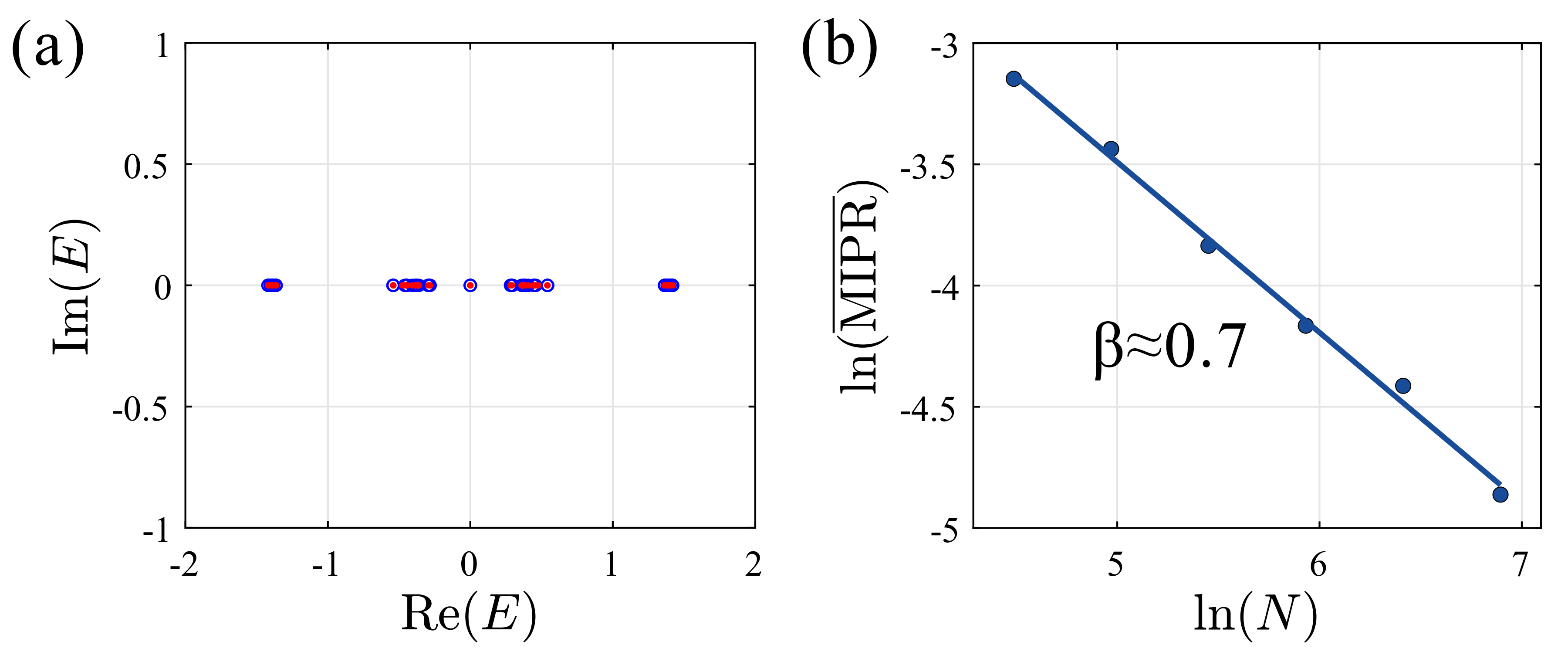}
    \caption{(a) Complex energy spectrum in the complex plane for PBC (blue circles) and OBC (red points). Parameters are $N=89$, $J=1$, and $\theta=0$.
    (b) Finite-size scaling of the $\overline{\mathrm{MIPR}}$ versus  system size $N$ on a logarithmic scale. The solid line represent power-law fits. The statistical average is performed over $10^4$ random realizations of the global phase $\theta$. }
    \label{S1}
\end{figure}
Let us consider the Hermitian quasiperiodic HN model. The Hamiltonian of the system reads
\begin{equation}
\hat{H}=\sum_{n=1}^{N-1}\left(J_{n}^{R} \hat{c}_{n+1}^{\dagger} \hat{c}_{n}+J_{n}^{L} \hat{c}_{n}^{\dagger} \hat{c}_{n+1}\right)+\hat{H}_{B},
\label{Hermitian quasiperiodic HN model}
\end{equation}
with the hopping amplitudes $J_{n}^{L}=J_{n}^{R}=J \cos [2 \pi \alpha(n+1 / 2)+\theta]$. In this model, the OBC and PBC spectra are almost identical as shown in Fig.~\ref{S1}(a), indicating that the NHSE is no longer present. We observe the conventional quasiperiodic criticality in this model. This behavior is clearly demonstrated in Fig~\ref{S1}(b). 
To investigate the critical properties in this model, we analyze the mean inverse participation ratio (MIPR) for normalized eigenstates: $\operatorname{MIPR} = \frac{1}{N} \sum_{j=1}^{N} \left( \sum_{n=1}^{N}  \left|\psi_{n}^{(j)}\right|^{4}  \right)$.
We calculate the $\mathrm{MIPR}$ under PBC, and then a statistical average $\overline{\mathrm{MIPR}}$ is made for $10^4$ random realizations of the global phase $\theta$. As shown in Fig.~\ref{S1}(b), the $\overline{\mathrm{MIPR}}$ decreases as the system size $N$ increases, indicating that the critical eigenstates. For conventional quasiperiodic criticality, $I_n$ would be almost uniform along the sample, indicating the nature of the conventional criticality.

\section{Inverse participation ratio of Hatano-Nelson model}
\label{IPRofHN}
\begin{figure}[t]
    \centering
    \includegraphics[width=1\linewidth]{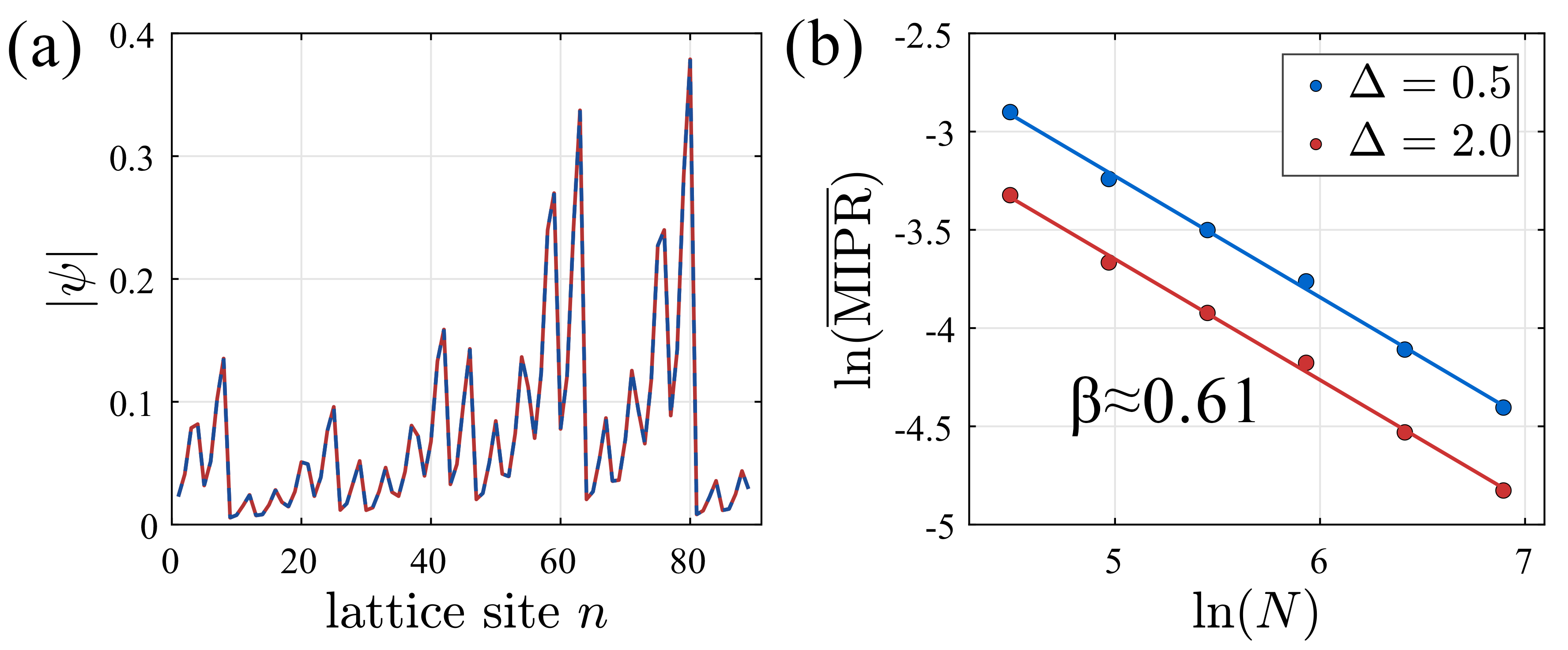}
    \caption{(a) Comparison of the spatial eigenstate profile $|\psi|$ for $N=89$ and $\theta=0$. The numerical diagonalization result (red dashed line) shows excellent agreement with the exact analytical solution (blue dashed line). 
    (b) Scaling analysis of the $\overline{\mathrm{MIPR}}$ versus system size $N$ on a logarithmic scale for the asymmetric case ($J=1, t=3$). Data for distinct interchain couplings are shown in blue ($\Delta = 0.5$) and red ($\Delta = 2.0$). The solid lines represent power-law fits. The statistical average is performed over $10^3$ random realizations of the global phase $\theta$.}
    \label{S2}
\end{figure}

In this section, we calculate the $\operatorname{IPR}$ of the eigenstates $\psi_{n}$ for the non-Hermitian quasiperiodic HN model. Since the energy spectrum and eigenstates are highly sensitive to boundary conditions due to the NHSE, we explicitly consider the PBC where the QNHSC arise. The PBC satisfies:
\begin{equation}
\psi_{n+N} = \psi_{n},\quad J_{n+N}^{R,L}=J_{n}^{R,L},
\label{Bian_Jie_Tiao_Jian}
\end{equation}
The $\operatorname{IPR}$ is defined as:
\begin{equation}
\operatorname{IPR} \equiv \frac{\sum_{n=1}^{N} \lvert \psi_{n} \rvert^{4}}{\left(\sum_{n=1}^{N} \lvert \psi_{n} \rvert^{2}\right)^{2}}.
\end{equation}
Upon substituting the nonunitary gauge transformation given by Eq.~\eqref{Nonunitary_Gauge_Transformation} into the IPR definition, we obtain:
\begin{equation}
\operatorname{IPR}=\frac{\sum_{n=1}^{N}\left|\phi_{n}\right|^{4} \exp \left(4 \Gamma_{n}\right)}{\left(\sum_{n=1}^{N}\left|\phi_{n}\right|^{2} \exp \left(2 \Gamma_{n}\right)\right)^{2}}.
\label{IPR}
\end{equation}
The transformed amplitudes $\phi_n$ satisfy the disorder-free difference equation $E \phi_{n}= \phi_{n+1}+\phi_{n-1}$. Deriving from Eqs.~\eqref{Bian_Jie_Tiao_Jian} and \eqref{Nonunitary_Gauge_Transformation}, the transformed field satisfies the boundary condition:
\begin{equation}
\phi_{N+1} = \phi_1 e^{-\Gamma_{N+1}}.
\end{equation}
The solution is given by the Bloch-like ansatz:
\begin{equation}
\phi_n = \exp\left( i q n - \frac{n}{N}\Gamma_{N+1} \right), \quad q = \frac{2\pi m}{N},
\label{spectral_problem_solution}
\end{equation}
with $m = 0,1,\dots,N-1$. A key observation is that the modulus $|\phi_n|$ is independent of the momentum $q$:
\begin{equation}
|\phi_n| = \exp\left( -\frac{n}{N}\Gamma_{N+1} \right).
\end{equation}
By combining the nonunitary gauge transformation with the solution of Eq.~\eqref{spectral_problem_solution}, we obtain the exact analytical form of the eigenstates: $\left| \psi_{n} \right| \propto \exp\left\{\Gamma_{n}-\frac{n}{N} \Gamma_{N+1}\right\}$. To validate this analytical result, we compare in Fig.~\ref{S2}(a) the eigenstates obtained from numerical diagonalization with the exact analytical prediction. The numerical result (red dashed line) is in perfect agreement with the analytical formula (blue dashed line) for a system size $N=89$ at $\theta=0$, confirming the rigorousness of the exact solution. Substituting this result back into Eq.~\eqref{IPR}, we obtain the exact closed-form expression:
\begin{equation}
\mathrm{IPR}=\frac{\sum_{n=1}^{N} \exp \left(4 \Gamma_{n}-4 \Gamma_{N+1} \frac{n}{N}\right)}{\left[\sum_{n=1}^{N} \exp \left(2 \Gamma_{n}-2 \Gamma_{N+1} \frac{n}{N}\right)\right]^{2}}.
\end{equation}
This derivation explicitly demonstrates that the $\operatorname{IPR}$ is identical for all eigenstates regardless of their eigenenergy $E$. While energy-independent, the IPR remains sensitive to the specific realization of the global phase $\theta$ via $\Gamma_n$.

\section{Inverse participation ratio of Hatano-Nelson ladder model}
\label{IPRofHNL}

In this section, we derive the exact analytical expression for the IPR of the eigenstates in the HN ladder model. We focus on the symmetric case $J=t$, where the restoration of universality allows for an exact solution.
As discussed in the main text, the system under PBC satisfies:
\begin{equation}
\psi_{n+N}^{A,B} = \psi_{n}^{A,B},\quad J_{n+N}^{R,L}=J_{n}^{R,L},\quad T_{n+N}^{R,L}=T_{n}^{R,L},
\end{equation}
where $N$ is the number of unit cells. We apply the sublattice-dependent nonunitary gauge transformation:
\begin{equation}
\begin{aligned}
\psi_{n}^\mathrm{A} = \phi_{n}^\mathrm{A} e^{\Gamma_n},\quad
\psi_{n}^\mathrm{B} = \phi_{n}^\mathrm{B} e^{\Gamma_n},
\label{nonunitary gauge transformation2}
\end{aligned}
\end{equation}
where $\Gamma_n$ is defined as in the single-band case (with $J$ replacing $t$ for the second chain since $J=t$). The IPR for the two-band system is defined as:
\begin{equation}
\mathrm{IPR} = \frac{\sum_{n=1}^{N} \left( |\psi_{n}^{A}|^4 + |\psi_{n}^{B}|^4 \right)}{\left[ \sum_{n=1}^{N} \left( |\psi_{n}^{A}|^2 + |\psi_{n}^{B}|^2 \right) \right]^2}.
\end{equation}
Substituting the transformation, this becomes:
\begin{equation}
\mathrm{IPR} =\frac{\sum_{n=1}^{N} \left( \left| \phi_{n}^{A} \right|^{4} + \left| \phi_{n}^{B} \right|^{4} \right)e^{4 \Gamma_{n}}}{\left[\sum_{n=1}^{N} \left( \left| \phi_{n}^{A} \right|^{2} + \left| \phi_{n}^{B} \right|^{2} \right)e^{2 \Gamma_{n}}\right]^{2}}.
\label{eq:IPR_2band_inter}
\end{equation}
In the transformed frame with $J=t$, the amplitudes satisfy the coupled difference equations:
\begin{equation}
\begin{aligned}
E\phi_{n}^{A} &= \phi_{n-1}^{A} + \phi_{n+1}^{A} + \Delta\phi_{n}^{B}, \\
E\phi_{n}^{B} &= \phi_{n-1}^{B} + \phi_{n+1}^{B} + \Delta\phi_{n}^{A}.
\end{aligned}
\end{equation}
These equations can be decoupled by introducing the symmetric and antisymmetric superpositions $\phi_{n}^{\pm} = (\phi_{n}^{A} \pm \phi_{n}^{B})/\sqrt{2}$. Both states $\phi_{n}^{\pm}$ satisfy a simple disorder-free hopping equation with shifted energies $E \mp \Delta$. Consequently, the solutions are Bloch states with uniform moduli $|\phi_{n}^{A}| = |\phi_{n}^{B}| = |\phi|$. The twisted boundary condition imposed by PBC leads to the solution:
\begin{equation}
|\phi_{n}^{A}| = |\phi_{n}^{B}| = \exp\left( -\frac{n}{N} \Gamma_{N+1} \right),
\end{equation}
where the momentum dependence $e^{iqn}$ (with $q=2\pi m/N$) disappears upon taking the modulus.
Substituting $|\phi_{n}^{A}| = |\phi_{n}^{B}|$ into Eq.~\eqref{eq:IPR_2band_inter}, we obtain:
\begin{equation}
\operatorname{IPR}=\frac{\sum_{n=1}^{N} \exp \left(4 \Gamma_{n}-4 \Gamma_{N+1} \frac{n}{N}\right)}{2\left(\sum_{n=1}^{N} \exp \left(2 \Gamma_{n}-2 \Gamma_{N+1} \frac{n}{N}\right)\right)^{2}}.
\end{equation}
This matches Eq.~\eqref{IPR_2band} in the main text. It confirms that for $J=t$, the IPR is independent of the eigenenergy $E$ and is solely determined by the global phase $\theta$ via the $\Gamma_n$.

When the hopping amplitudes become asymmetric ($J\neq t$), the exact symmetry leading to energy-independent eigenstate profiles is broken. Mathematically, the nonunitary gauge transformation induces spatially dependent effective coupling, which acts as quasiperiodic potentials. Consequently, the IPR becomes dependent on the eigenenergy $E$ and is sensitive to the inter-chain coupling strength $\Delta$, which controls the hybridization between the sublattices.

To investigate the critical properties in this regime, we analyze the $\overline{\mathrm{MIPR}}$. Figure~\ref{S2}(b) displays the scaling behavior of the $\overline{\mathrm{MIPR}}$ versus system size $N$ for the asymmetric case with $J=1$ and $t=3$. We compare two distinct coupling strengths: $\Delta = 0.5$ (blue points) and $\Delta = 2.0$ (red points). The solid lines represent linear fits on a logarithmic scale, revealing a clear power-law decay $\overline{\operatorname{MIPR}} \sim N^{-\beta}$. Remarkably, despite the breaking of the exact solvability condition, the scaling analysis yields a fractal dimension $\beta \approx 0.61$. This finding suggests that the quasiperiodic skin criticality and its multifractal universality class are robust against perturbations to the hopping symmetry.

\bibliography{reference}
\end{document}